# STUDIES OF A TERAWATT X-RAY FREE-ELECTRON LASER

H.P. Freund[1,2] and P.J.M. van der Slot[3]

[1]Department of Electrical and Computer Engineering, University of New Mexico, Albuquerque, New Mexico USA
[2]NoVa Physical Science and Simulations, Vienna, Virginia, USA
[3]Mesa+ Institute for Nanotechnology, Department of Science and Technology, University of Twente, Enschede, the Netherlands

The possibility of terawatt (TW) x-ray free-electron lasers (XFELs) has been discussed using novel superconducting helical undulators. In this paper, we consider the conditions necessary for achieving powers in excess of 1 TW in a 1.5 Å FEL. Using h the MINERVA simulation code, an extensive steady-state analysis has been conducted using a variety of undulator and focusing configurations. In particular, strong focusing using FODO lattices is compared with the natural, weak focusing inherent in helical undulators. It is found that the most important requirement to reach TW powers is extreme transverse compression of the electron beam in a strong FODO lattice in conjunction with a tapered undulator. We find that when the current density reaches extremely high levels, that characteristic growth length in the tapered undulator becomes shorter than the Rayleigh range giving rise to optical guiding. We also show that planar undulators can reach near-TW power levels. In addition, preliminary time-dependent simulations are also discussed and show that TW power levels can be achieved both for the self-seeded MOPA and pure SASE.

## I. INTRODUCTION

The number of x-ray Free-Electron Lasers (XFELs) is increasing around the world [1-4] and the user community for these light sources is growing as well. Along with this growth of the user community, we confidently expect that novel and important new applications will be found. The first operational XFEL, the Linac Coherent Light Source (LCLS) at the Stanford Linear Accelerator Center [1], produces approximately 20 GW pulses of 1.5 Å photons at a repetition rate of 120 Hz, and the other x-ray FELs produce similar power levels. In order to be proactive and produce higher photon fluxes for the rapidly developing user community, research is under way around the world in techniques to produce still higher peak powers.

Recent simulations [5] based on the GENESIS simulation code [6] and using a novel super-conducting helical undulator design and a quadrupole FODO lattice indicate that a terawatt (TW) XFEL is possible. The configuration studied was based upon a self-seeding scheme whereby the electron beam is propagated through an undulator long enough to achieve exponential gain via Self-Amplified Spontaneous Emission (SASE) but not long enough to reach saturation. The SASE radiation produced is then passed through a monochromator after which the filtered optical pulse is reintroduced to the electron beam in a longer undulator. This seed is assumed to be at power levels far in excess of the noise; hence, the subsequent interaction is equivalent to that of a seeded XFEL. The seed power is assumed to be at MW power levels, and TW output power levels are found after an additional 100 m of a tapered undulator. Both steady-state and time-dependent simulations were described where it was found, as expected, that the power was reduced when time-dependence was included in the simulations.

This result represents an enhancement of the output power in the tapered undulator configuration of nearly two orders of magnitude over the saturated power in a uniform undulator. It has long been recognized that efficiency enhancements are possible in FELs using a tapered undulator [7-9]. Historically, however, experiments have shown efficiency enhancements using a tapered undulator of factors of 3 – 5 [10-12]. It is important, therefore, to understand what gives rise to such extreme efficiency enhancements with tapered undulators.

The determination of the optimal taper profile has been addressed in numerous papers for a variety of configurations [2,13-16], and neither a linear nor quadratic taper may be optimal for every configuration. Indeed, the optimal taper profile found in ref. 2 varied with distance along the undulator by a power of 2.1. A universal scaling law for the optimal taper profile was developed [13] in which it was demonstrated that the optimal profile should be quadratic for the case of a wide electron beam but linear for a thin beam. Within the context of this analysis, a wide beam is defined over a length where the Fresnel number $N \geq 1$ while the thin beam is characterized by $N \ll 1$, where the Fresnel number is defined as $N = 2\pi\sigma^2/\lambda z$ for an rms beam size $\sigma$ and free space wavelength $\lambda$ over an undulator length $z$. For the cases considered here, this implies that the thin beam limit is applicable for undulators longer than about 445 m while the wide beam is applicable for undulator lengths up to about 4.5 – 18.5 m. Since the undulators under consideration fall between these limits, simple linear or quadratic taper profiles are only approximations. However, more complex tapering schemes have also been discussed where the specific variation of the undulator with distance is optimized on an undulator-by-undulator basis in both simulation [15] and in the laboratory [16] depending on the detailed phase space evolution of the electron beam. In such an optimization scheme, the detailed taper profile may not follow a simple power law and may, indeed, not even show a monotonic variation in the undulator field. A detailed study of the optimal taper profile as a function of transverse beam size, therefore, is a fertile field for future research and will be reported in a forthcoming publication.

In view of the complex issues associated with determining the optimal taper profile, our purpose in this paper is to study the fundamental requirements for achieving TW power levels in XFELs rather than finding an absolute optimum. To this end, we consider various configurations necessary to achieve TW power levels in an XFEL based upon a linear taper profile in conjunction with a strong-focusing FODO lattice using both helical and planar undulators. Simulations are described using the MINERVA simulation code [17]. This represents a preliminary analysis using both steady-state and time-dependent simulations. The principal result we find is that near-TW or TW power levels can be obtained using extreme focusing of the electron beam by the FODO lattice with either a helical or planar tapered undulator. Hence, it is the extremely tight focusing imposed by the FODO lattice that gives rise to such extreme efficiency enhancements. For the cases considered, the rms electron beam radius can be as small as 7 – 8 μm and the peak current densities reach 25 – 30 GA/cm$^2$. In contrast to expectations, we find that optical guiding occurs even for a tapered undulators in the limit of such extreme focusing when the characteristic growth length in the tapered undulator is shorter than the Rayleigh range. Simulations are performed using both long, continuous undulators and segmented undulators, and the taper is optimized with respect to both the start-taper point and the taper slope. This technique can be used for pure SASE XFELs as well as schemes where a monochromator is used to selectively narrow the SASE linewidth.

The organization of the paper is as follows. A brief description of the MINERVA simulation code is given in Sec. II. Steady-state simulations are discussed in Sec. III under the assumption of self-seeding. This is effectively a Master Oscillator Power Amplifier (MOPA). Simulations of a long, single-section helical undulator are described. We compare the performance of a weak-focusing helical undulator with that for a helical undulator with a strong-focusing FODO lattice. While a single, 100 m long undulator is not practical, this configuration serves to illustrate the effects of varying focusing strengths on the performance of a tapered undulator. We also compare the performance of the single, long undulator with that of a segmented undulator with identical period and field strength. Simulations are discussed for a planar undulator system with strong-focusing showing similar increases in the output power when extreme focusing of the electron beam is applied. Following the steady-state simulations, preliminary time-dependent simulations are also discussed for both the self-seeded and SASE configurations in Sec. IV. While the previous simulations were performed under the assumption that the electron beam had a Gaussian transverse profile, Emma *et al*. [5] found that improved performance was obtained using parabolic and flat-top transverse profiles. In order to verify this result, we describe the performance found using a parabolic transverse profile in Sec. V. A summary and discussion is given in Sec. VI.

**II. THE MINERVA SIMULATION CODE**

The MINERVA simulation code [17] is based on a three-dimensional, time-dependent nonlinear formulation of the interaction that is capable of modeling a large variety of FELs including amplifier, oscillator, and self-amplified spontaneous emission (SASE) configurations.

MINERVA employs the Slowly-Varying Envelope Approximation (SVEA) in which the optical field is represented by a slowly-varying amplitude and phase in addition to a rapid sinusoidal oscillation. The optical field is described by a superposition of Gaussian modes. The field equations are then averaged over the rapid sinusoidal time scale and, thereby, reduced to equations describing the evolution of the slowly-varying amplitude and phase. Time-dependence is treated using a breakdown of the electron bunch and the optical pulse into temporal *slices* each of which is one wave period in duration. The optical *slices* slip ahead of the electron slices at the rate of one wavelength per undulator period. MINERVA integrates each electron and optical *slice* from $z \rightarrow z + \Delta z$ and the appropriate amount of slippage can be applied after each step or after an arbitrary number of steps by interpolation.

Particle dynamics are treated using the full Newton-Lorentz force equations to track the particles through the optical and magnetostatic fields. The formulation tracks the particles and fields as they propagate along the undulator line from the start-up through the (linear) exponential growth regime and into the nonlinear post-saturation state. MINERVA includes three-dimensional descriptions of linearly polarized, helically polarized, and elliptically polarized undulators including the fringing fields associated with the entry/exit transition regions. This includes an analytical model of an APPLE-II undulator. Additional magnetostatic field models for quadrupoles and dipoles are also included. These magnetic field elements can be placed in arbitrary sequences to specify a variety of different transport lines. As such, we can set up field configurations for single or multiple wiggler segments with quadrupoles either placed between the undulators or superimposed upon the undulators to create a FODO lattice. Dipole chicanes can also be placed between the undulators to model various optical klystron and/or high-gain harmonic generation (HGHG) configurations. The fields can also be imported from a field map. It is important to remark that the use of the full Newton-Lorentz orbit analysis allows MINERVA to treat self-consistently both the entry/exit taper regions of undulators, and the generation of harmonics of the fundamental resonance.

In order to apply the formulation to the simulation of FEL oscillators, an interface has been written between MINERVA and the optical propagation code OPC [18,19]. Oscillator simulations proceed by tracking the output optical pulse from the undulator as simulated by MINERVA, through the resonator and back to the undulator entrance using OPC, after which the optical field is then imported into MINERVA for another pass through the undulator. This process is repeated for as many passes through the undulator and resonator as required for the oscillator to achieve a steady-state.

The unique features/advantages of MINERVA can be summarized as follows. Since electron dynamics are integrated using the complete Newton-Lorentz equations, MINERVA treats the complete electronic interaction with the magnetostatic and electromagnetic fields. This permits the simulation of the entry/exit transitions from the undulators, quadrupoles and dipoles; hence, the actual locations, lengths, and field strengths are used. In addition, all harmonic elements of the trajectories are included self-consistently so that harmonic generation is implicitly included in the formulation. Finally, since the optical field is described by a superposition of Gaussian modes, the dynamical equations include the evolution equations of both the electrons and the field amplitudes. Because of this, both the electrons and the fields are propagated self-consistently through the gaps between the undulators so that the relative phase advance between the electrons and the fields in these regions is tracked self-consistently.

### III. STEADY-STATE SIMULATIONS

We first discuss steady-state simulations This permits rapid scans over a large variety of configurations while capturing the essential underlying physics. In particular, many simulation runs are needed to optimize a tapered undulator configuration with respect to the start-taper point and the taper slope. However, we expect that the slippage of the optical field relative to the electrons will result in some degradation of the interaction and time-dependent simulations are required to accurately describe an actual experimental configuration. To this end, we also discuss some initial time-dependent simulations in Sec. IV.

#### A. The Case of a Helical Undulator

The configuration that we consider is based upon a self-seeding scenario [20] in which the interaction in a SASE FEL is halted at an early stage and then passing the optical field through a monochromator to extract a narrow band of the SASE radiation after which this narrow bandwidth light is then re-injected into the undulator in synchronism with the electron beam. Hence, the light acts as the seed for the amplifier section in a Master Oscillator Power Amplifier (MOPA).

The electron beam is assumed to be characterized by an energy of 13.22 GeV, a peak current of 4000 A, an rms energy spread of 0.01% and normalized emittances of 0.3 mm-mrad in both the $x$- and $y$-directions. Following Emma *et al.* [5] this corresponds to the simulation of a fresh bunch [21] in the MOPA section. The transverse profile of the electron beam is assumed to be characterized by a Gaussian distribution, and the beam is matched into either the natural focusing of a helical undulator or the FODO lattice/helical undulator system.

We initially study the case of a single, long undulator with a period of 2.0 cm and a peak on-axis amplitude of 16.1 kG. This is equivalent to an undulator parameter $K$ = 3.01, and yields a resonance at a wavelength of 1.5 Å. The simulations do not include the initial SASE region prior to the monochromator, and the undulator in the MOPA region is assumed to be 100 m in length.

While it is not practical to construct an undulator that is 100 m in length, this model is useful to study the essential physics of the interaction in a tapered undulator with strong focusing applied. Since (1) optical guiding ceases in the gaps between undulator segments and (2) the optimal phase match between undulators changes as the field strength decreases, it is expected that the interaction will be less efficient in a segmented undulator. Nevertheless, an equivalent simulation of a segmented undulator without making allowance for the changing phase match, will be described for comparison.

| $L_{FODO}$ (m) | Gradient (kG/cm) | $(\beta_x^2 +\beta_y^2)^{1/2}$ (m) | Radius (μm) | $x_{rms}/y_{rms}$ (μm) |
|---|---|---|---|---|
| 2.2 | 26.40 | 6.49 | 10.3 | 8.03/6.22 |
| 3.3 | 18.69 | 9.61 | 12.2 | 9.68/7.47 |
| 4.4 | 14.02 | 12.32 | 14.0 | 11.31/8.72 |
| 5.0 | 12.34 | 14.57 | 15.2 | 12.06/9.29 |
| 5.5 | 11.22 | 16.04 | 16.2 | 12.65/9.74 |
| 6.0 | 10.28 | 17.49 | 16.5 | 13.0/10.2 |
| 6.6 | 9.35 | 19.26 | 16.9 | 13.9/10.7 |
| 8.8 | 7.01 | 25.69 | 19.2 | 16.0/12.3 |
| 11.1 | 5.56 | 32.41 | 21.0 | 18.0/13.8 |

Table 1 FODO lattice parameters.

In order to study the effect of increasingly strong focusing, we consider nine different FODO lattices as shown in Table 1. In each case the quadrupole length is assumed to be 0.074 m. The leftmost column in the table represents the length of the FODO cell while the second column is the field gradient. The third column is a measure of the average $\beta$-function, while the two rightmost columns describe the rms beam radius and the initial beam sizes in the $x$- and $y$-directions. The Twiss-$\alpha$ parameters are not shown but are $\alpha_x \approx 1.3$ and $\alpha_y \approx -0.77$.

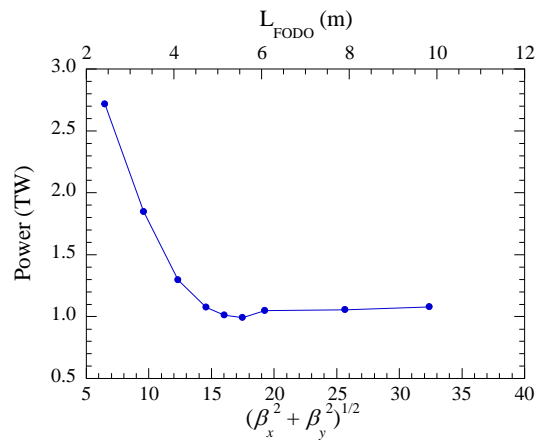

Fig. 1 Optimized output power after 100 m for the nine FODO lattices.

A summary showing the optimized output power for a linear taper profile after 100 m for the different FODO lattices is shown in Fig. 1 for a seed power of 5 MW. However, there appear to be two regimes associated with strong-focusing. At the longer FODO cell lengths ($L_{FODO}$ = 6.6 m, 8.8 m, and 11.1 m), MINERVA predicts output powers of approximately 1.1 TW. However, there is a more dramatic increase in the output powers as the FODO cell length decreases below 6.0 m. In this regime, we observe an approximately linear increase in the output power with decreasing $\beta$-functions, and where the maximum output power exceeds 2.7 TW for the cases under consideration.

The current density increases as the FODO cell length decreases so that the Pierce parameter (Fig. 2) increases with decreasing $\beta$-functions, and this leads to increasing interaction strengths. The reason for the dramatic increase in the output power when the FODO cell length decreases below 6.0 m is twofold. In the first place, the stronger interaction strength yields optical guiding even in the tapered regime. In the second place, the smaller beam size results in a more coherent interaction with the optical field and this results in a higher trapping fraction.

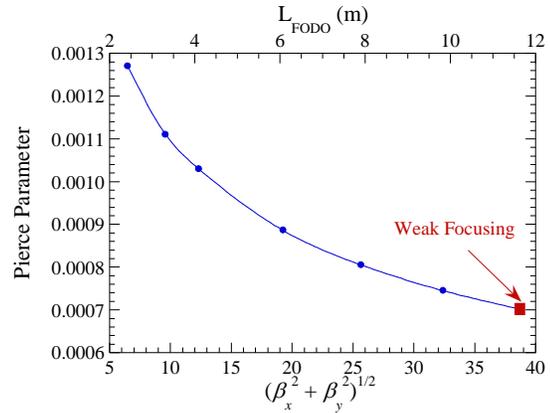

Fig. 2 The variation in the Pierce parameter with the $\beta$-function.

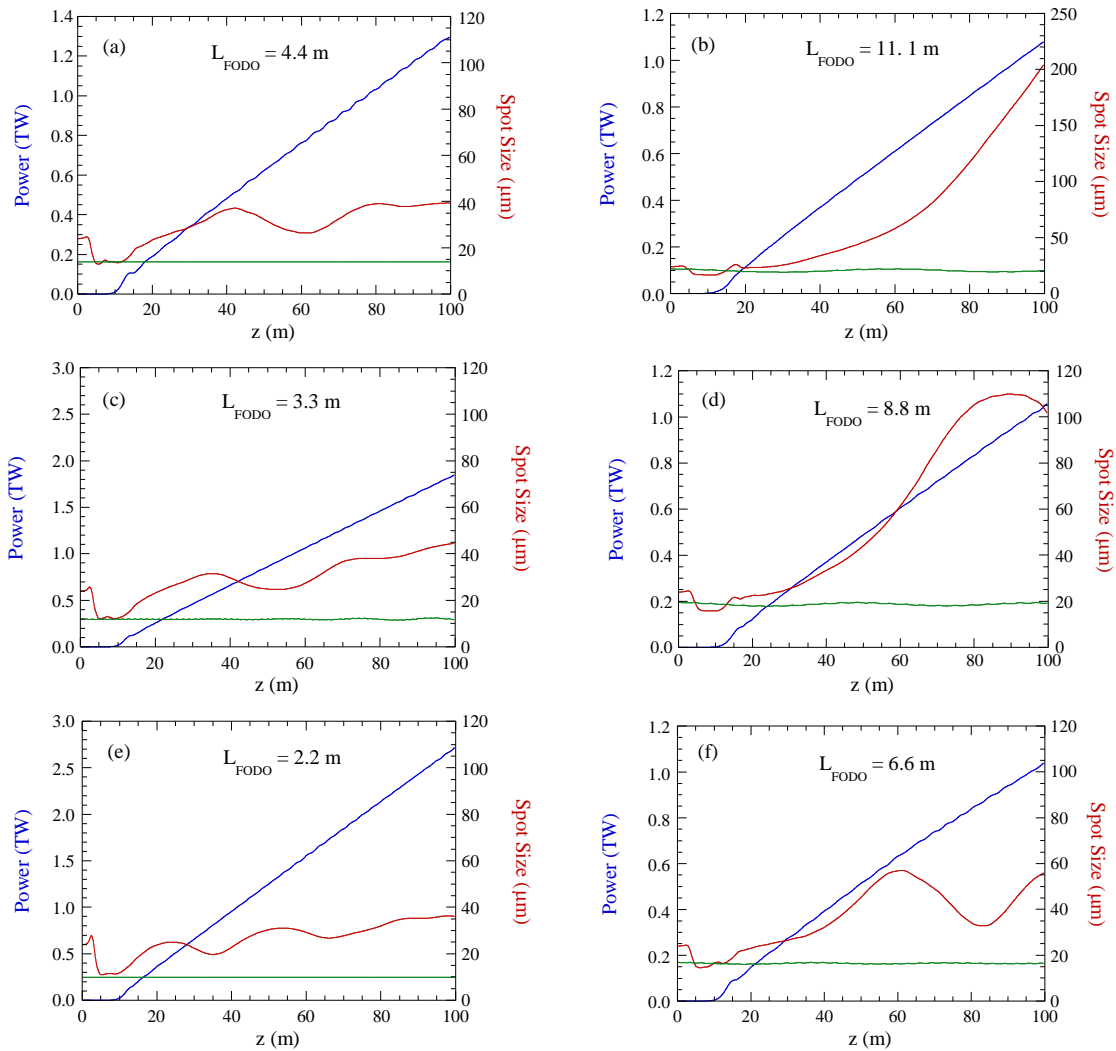

Fig. 3 Plots of the power (left axis, blue) and spot size (right axis, red) versus distance for different levels of focusing. The green line is the rms electron beam radius. Observe that the optical field is guided when the length of the FODO cell is less than about 4.4 m and is transitional with lesser guiding when the FODO cell length is 6.6 m.

The evolution of the output power (left axis, blue) and the spot size of the optical field (right axis, red) versus position in the undulator is shown in Fig. 3 for six choices of the FODO lattice. The green line represents the rms electron beam radius. In all of these cases, saturation in the uniform undulator occurs after between 9 – 12 m, and the optimal start-taper points range from about 9 – 12 m. Figures 4b, 4d, and 4f correspond to the longest FODO cells with the lower power levels. It is evident from these figures that substantial diffraction occurs after the start of the taper where the optical field expands from about 10 – 20 μm at the start-taper point to between 60 – 200 μm after 100 m. This is in stark contrast with what is found when the FODO cell is shorter than 6.0 m in length (shown in Figs. 4a, 4c, and 4e). As shown in the figure, the optical mode experiences substantial guiding in the tapered regime and the maximum expansion found after 100 m of undulator is to a mode radius of about 40 μm.

Optical guiding [22] is thought to be composed of two effects: gain guiding and refractive guiding. Gain guiding occurs because the amplification of the field occurs only within the electron beam. In the exponential regime, theory shows that the interaction leads to a complex solution for the wavenumber which has a shift in the real part of the solution that is proportional to the growth rate and which shifts the refractive index and causes the refractive guiding. However, these two effects cannot be separated and optical guiding occurs when the characteristic growth length of the optical field is shorter than the Rayleigh range. This is usually the case in the exponential growth regime, but, it is not typically found in a tapered undulator where the power grows more slowly than exponential. However, the extreme focusing that we find when $L_{FODO} \leq 6.0$ m leads to sufficiently rapid growth that optical guiding can occur. This is seen by noting that the power grows linearly in the linear tapered region as $P(z) = P_0[1 + (z - z_0)/L_G]$, where $L_G$ is the characteristic growth length. When $L_{FODO} = 2.2$ m (see Fig. 3e), the characteristic growth length is $L_G \approx 12.3$ m and the Rayleigh range is about 3 m at the start of the tapered region. Since the Rayleigh range is shorter than the gain length, diffraction initially dominates over the amplification and the optical mode expands. The spot size, and the Rayleigh range, grows until about the 20 m point where the Rayleigh range is about 11.8 m, which is comparable to the growth length, after which the Rayleigh range increases further and the optical field is guided and largely confined to within a spot size of about 25 – 40 μm. Observe that the spot size exhibits bounded oscillations indicating that the guiding is modulated by the detailed phase space evolution of the electrons which undergo synchrotron oscillations in the ponderomotive potential with a period in the range of 20 – 30 m.

Similar behavior is found in all cases where $L_{FODO}$ is less than about 6.0 m, while no optical guiding is found when $L_{FODO} > 6.6$ m. The range 5.0 m < $L_{FODO}$ < 6.6 m is transitional between these two regimes.

It was pointed out by Jiao *et al*. [22], that the "decreasing of refractive guiding is the major cause of the efficiency reduction, particle detrapping, and then saturation of the radiation power" in a tapered undulator. Hence, the continuation of optical guiding, and the associated increase in the trapping fraction, in the case of extreme transverse compression of the electron beam is an important factor in reaching TW power levels.

It should also be remarked that the interaction might be further optimized by varying the electron beam size within the tapered undulator [23]. Simulation of this requires a multi-parameter optimization of the FODO lattice including both the quadrupole field gradients and positions. Given the length of the FODO lattice and the number of undulators and quadrupoles, this is an extremely arduous task, and will be deferred to a future publication.

The physics of optical guiding in the a tapered undulator has been studied analytically [24] in which it was shown that refractive guiding varies inversely with the on-axis field strength which typically reaches an asymptotic limit. Once that limit is achieved, refractive guiding ceases and diffraction takes hold leading to expansion of the optical field. At the same time, the optical power continues to increase linearly. This is, effectively, what is shown in Figs. 4b, and 4d. It is important to observe that the results shown in Figs. 4a, 4c, 4e, and to some extent in Fig. 4f, represent a new regime where, while the power still increases linearly with distance, it grows rapidly enough to overcome diffraction; hence, optical guiding becomes an important effect.

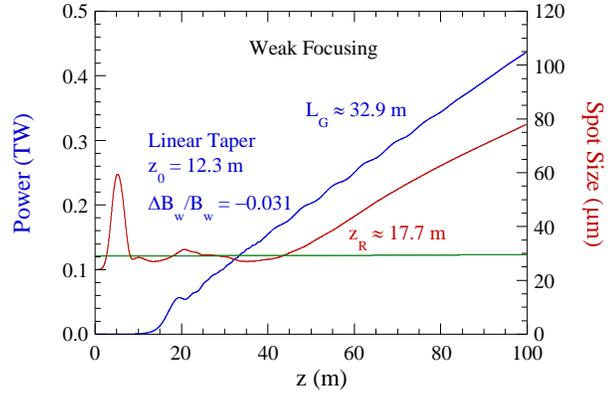

Fig. 4 Plot of the evolution of the optimized power and the spot size versus distance for a weak-focusing helical undulator.

For the purpose of comparison, the case of a weak-focusing helical undulator is shown in Fig. 4 where we plot the power (left axis, blue) and spot size (right axis, red) versus distance along the undulator, and where the green line represents the rms electron beam radius. Apart from the choice of natural focusing (*i.e*., without quadrupoles), all the beam and undulator parameters are the same as used for the strong focusing cases except that the $\beta$-function in this case is 38.7 m. In contrast, the longest $\beta$-function for the strong focusing examples was 32.4 m. Saturation in a uniform undulator is found after about 14 – 15 m and the optimal taper starts after 12.3 m with a total down-taper of 3.1% over the 100 m of undulator resulting in an output power of about 0.43 TW. The growth length $L_G \approx 32.9$ m which is longer than the Rayleigh range of about 17.7 m for an optical field that has a spot size comparable to the rms electron beam radius. As a result, while some guiding persists after the end of the exponential gain region out to about 40 m of undulator, the optical field expands thereafter.

The second major factor in achieving such extreme enhancements of the efficiency is how strong-focusing affects the trapping fraction. In this regard, we find that the trapping fraction is highest at about 50% of the beam for the shortest FODO cell length ($L_{FODO}$ = 2.2 m) and decreases as the FODO cell length increases. This is shown in Fig. 5 where we plot variation in the trapping fraction after 100 m of undulator versus the $\beta$-function. This is related to the optical guiding where the mode size remains closest to that of the transverse extent of the electron beam and this enhances the coupling of the field to the electrons.

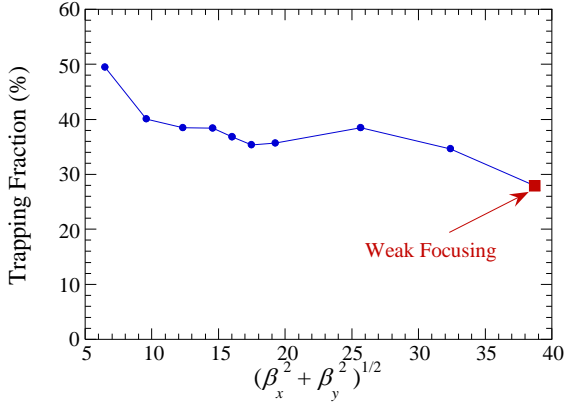

Fig. 5 Variation in the trapping fraction after 100 m with the $\beta$-function.

Examples of the spent beam distributions for (a) strong focusing with the highest power case ($L_{FODO}$ = 2.2 m), (b) strong focusing for the case of $L_{FODO}$ = 6.6 m where diffraction is still dominant, and (c) for the weak focusing case are shown in Fig. 6. It is clear from the figures that the extreme focusing has a significant impact on the trapping fraction, which is close to 50% for the strongest focusing lattice but only about 28% for the weak focusing undulator. In addition, a greater amount of energy has been extracted from the trapped portion of the beam when the strong focusing lattice is used.

We now consider a segmented helical undulator with the strong-focusing FODO lattice with $L_{FODO}$ = 2.2 m. In order to configure this undulator/FODO lattice properly, the undulators are 0.96 m in length with 46 periods (for a 2.0 cm period) with one period each in the entry and exit transition, and the gaps between the undulators are 0.16 m in length. Because MINERVA does not automatically select the optimum phase shift between the undulator sections, we have adjusted the wiggler field strength slightly to 16.135 kG in order to optimize the phase shift between undulator segments in the uniform undulator section for a resonance at 1.5 Å.

It should be remarked that since the undulator strength decreases in the tapered section, the factors controlling the optimal phase shift will also vary. The optimal phase shift can be selected by inserting phase shifters between the undulators or by varying the gap lengths. However, the gaps are quite short given the short length of the FODO cell and this would make it difficult to insert phase shifters in the configuration under consideration here. In addition, changing the gap lengths could, in principle, affect the lengths of the FODO cells. Of course, it may be that the interaction can be further optimized by changing the focus of the electron beam along the undulator which implies varying the parameters of the FODO lattice. As a result, it is clear that a more complete optimization of the segmented undulator configuration is a complicated procedure [22] which is beyond the scope of the present study. As such, therefore, we shall restrict the present analysis to an optimization over the start-taper segment and the (linear) taper profile, but it should be recognized that this is not a complete optimization.

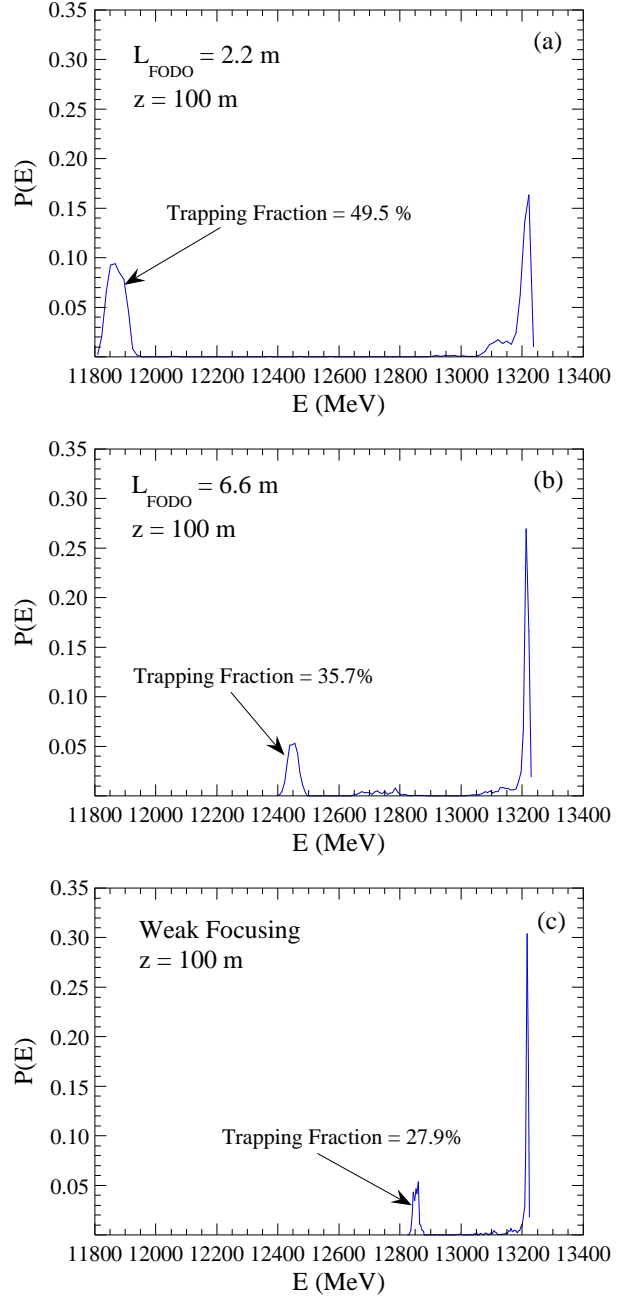

Fig. 6 The spent beam distributions for (a) strong focusing with $L_{FODO}$ = 2.2 m, (b) strong focusing with LFODO = 6.6 m, and (c) weak focusing.

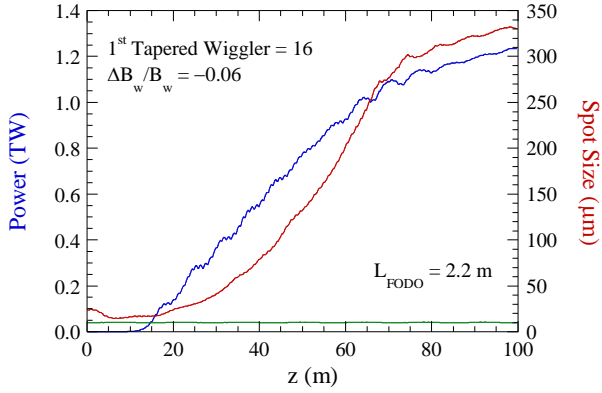

Fig. 7 Evolution of the power and spot size for an optimized tapered undulator line. Note this optimization is incomplete and includes only a single choice for the gap lengths separating the undulators.

The evolution of the power for an optimized taper profile and the corresponding spot size are shown in Fig. 7. The start-taper point is found to be the 16$^{th}$ undulator for a seed power of 5 MW, and the optimal (linear) down taper is 6%. As expected, the effect of the segmented undulator degrades the interaction because the optical field is not guided in the gaps and the phase shift may not be optimized, and this is found to be the case. The output power reaches approximately 1.25 TW which is reduced relative to the 2.7 TW found previously. Nevertheless, this still represents an enhancement by a factor of more than an order of magnitude over the saturated power in a uniform undulator.

It should be noted that the output power and spot size shown in Fig. 9 is comparable to that found for the single long undulator when optical guiding is absent (see Figs. 4b, 4d, and 4f). This is because (1) the power growth in the tapered region is reduced relative to that obtained in the case of a single, long undulator due to phase mis-matches in the gaps between the undulators, and (2) there is diffraction in the gaps between the undulators. In regard to (1), the optimal phase match between undulators in the tapered region will vary from gap to gap due to decreasing undulator amplitudes, and no effort has been made to find the (different) optimal gap lengths in the tapered section.

**B. The Case of a Planar Undulator**

We now consider a flat-pole-face planar undulator with a 3.0 cm period and an on-axis field magnitude of 12.49 kG. These parameters correspond to the period and magnitude of the undulator in the LCLS; however, we consider the case of a single, long undulator for the present study. The electron beam has an energy of 13.64 GeV, a peak current of 4000 A, a normalized emittance of 0.4 mm-mrad, and an rms energy spread of 0.01%. This configuration is resonant at a wavelength of 1.5 Å. The performance of the self-seeded MOPA configuration is studied using two FODO lattices. The FODO lattice used in the LCLS had a FODO cell length of approximately 7.3 m and a field gradient of 4.05 kG/cm. Note that the quadrupole length that we have been using (0.074 m) also corresponds to the quadrupoles used in the LCLS. We compare the performance of the self-seeded MOPA based upon this FODO lattice with the extreme-focusing lattice shown in Table 1 with a FODO cell length of 2.2 cm. A seed power of 5 MW is used to determine the performance of the self-seeded MOPA for both FODO lattices.

The Twiss parameters used to match the electron beam into the 7.3 m long FODO cell correspond to initial rms sizes of 22 μm in the $x$-direction and 19 μm in the $y$-direction with Twiss-$\alpha$ parameters of $\alpha_x = 1.1$ and $\alpha_y = -0.82$. Saturation for this lattice is found after about 25 m at a power level of close to 10 GW. The optimized linear down-taper is found to correspond to a start-taper point of 20 m with a total down-taper of 1.9% over the additional 80 m of tapered undulator. The evolution of the power and spot size for this optimized taper is shown in Fig. 8. The output power reaches 0.12 TW after 100 m of undulator, which represents an enhancement over the saturation power for the uniform undulator by a factor of 12.

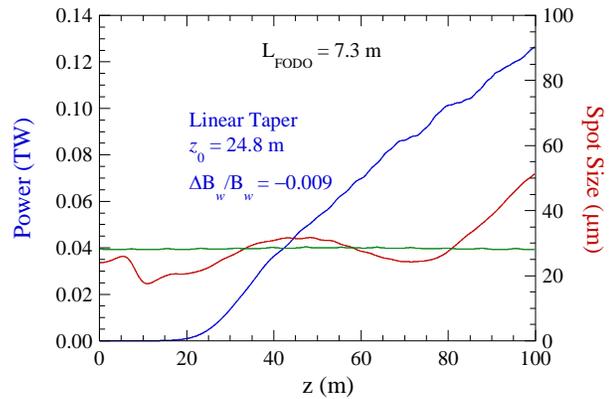

Fig. 8 Evolution of the power (left axis, blue) and spot size (right axis, red) for the optimized taper when $L_{FODO}$ = 7.3 m. The green line is the rms electron beam radius.

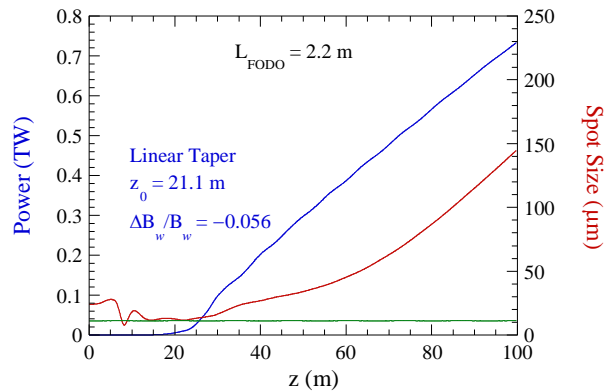

Fig. 9 Evolution of the power (left axis, blue) and spot size (right axis, red) for the optimized taper when $L_{FODO}$ = 2.2 m. The green line is the rms electron beam radius.

Electron beam propagation is determined largely by the FODO lattice rather than the undulator, and the initial Twiss parameters specified in Table 1 for the 2.2 m FODO cell are applicable for this planar undulator as well.

Saturation in the uniform undulator for this FODO lattice is found after about 23.5 m at a power level of 15 GW. The optimized linear taper profile is characterized by a start-taper point of 20.2 m and a total down-taper of 4.8% over the length of the taper region. The evolution of the power and spot size for the optimized taper profile is shown in Fig. 9, where the output power reaches 0.73 TW. This corresponds to an enhancement by a factor of about 50 over the saturated power in the uniform undulator, and a factor of six greater than the output power in the FODO lattice with the 7.3 m cell length.

The result shown in Fig. 9 refers to a planar undulator where we might expect a somewhat weaker interaction strength coming from the reduced *JJ*-factor; hence, the reduction in the output power and rate of growth in the tapered region relative to the helical undulator is not surprising. It is important to bear in mind, however, that while the spot size grows over the 80 m of tapered undulator, it only reaches about 145 microns, which is much less than found in Fig. 3b. As a result, this represents an intermediate regime where diffraction occurs more slowly and does not completely overwhelm the growth.

It is clear that the extreme-focusing FODO lattice will also bring the performance to near-TW power levels in 100 m long planar undulators.

## IV. TIME-DEPENDENT SIMULATIONS

We now describe preliminary time-dependent simulations based upon the single, long helical undulator configuration using the strongest FODO lattice with $L_{FODO}$ = 2.2 m. The basic electron beam parameters for these time-dependent simulations are the same as used in the steady-state simulations except that we assume a top-hat temporal profile for the electron bunch with a full-width duration of 24 fsec corresponding to a total bunch charge of 96 pC.

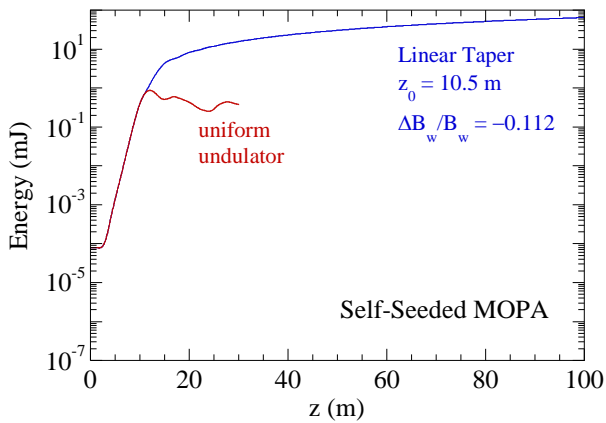

Fig. 10 Output pulse energies for the self-seeded MOPA configuration using uniform and tapered undulators.

A comparison of the output pulse energy found in simulation from the self-seeded MOPA for uniform and tapered undulators is shown in Fig. 10. Note that this assumes a peak seed pulse power of 5 MW yielding a pulse energy of 80 nJ. As shown in the figure, the saturated pulse energy in a uniform undulator is about 0.9 mJ. This pulse energy can be increased to about 65 mJ using a taper which starts at 10.5 m and is tapered downward by 11.2% over the remaining length of undulator. Estimation of the average output power can be obtained by dividing the pulse energy by the bunch duration and yields a peak output power of about 2.6 TW.

Turning to a time-dependent SASE simulation, a comparison of the evolution of the pulse energies for uniform and tapered undulators is shown in Fig. 11. As seen in the figure, the saturated pulse energy in the uniform undulator is about 1.1 mJ. This is increased to about 64 mJ using a downward linear taper starting at 14.0 m and extending over a total length of 100 m for a total down-taper of 10.8%. Estimation of the average power over the pulse yields a figure of about 2.6 TW.

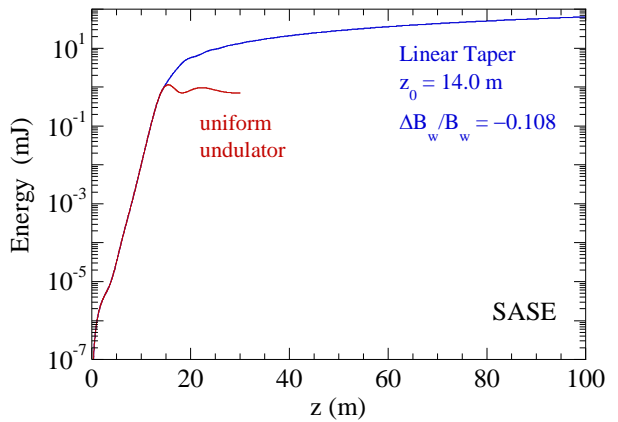

Fig. 11 Output pulse energies for the SASE configuration using uniform and tapered undulators.

There are two principal observations to be drawn from these time-dependent simulations. First, the average power of 2.6 TW is comparable to that found in the corresponding steady-state simulations. Second, the pulse energy found for pure SASE is comparable to that found for the self-seeded MOPA. Together, these observations imply that slippage is not an important effect. The slippage time over the course of the tapered undulator is $\tau_{slip} = N_w\lambda/c \approx 2.3$ fsec, where $N_w$ is the number of undulator periods in the tapered undulator, which is less than 10% of the bunch duration. Further, the ratio of the bunch duration to the slippage time yields the number of spikes in the SASE pulse. For this case, that implies that there are 10 – 11 spikes with durations of about 2.3 fsec. It is expected that the pulse energy for SASE should be less than for the self-seeded MOPA because, at least in part, the spiky nature of the SASE pulse causes de-trapping of the electrons as the spikes in the optical pulse slip relative to the electrons. However, the slippage time is comparable to the duration of the spikes, thereby minimizing the degradation expected for SASE for this set of parameters. Nevertheless, these are preliminary results that have not been exhaustively optimized, and a more complete optimization will be reported in a future publication.

## V. A PARABOLIC TRANSVERSE PROFILE

Simulations by Emma *et al*. [5] indicate that improved performance may be obtained using either a parabolic or flat-top transverse electron beam profile as compared with a Gaussian transverse profile. In order to evaluate the possible performance enhancement relative to the Gaussian transverse profile used above, we have optimized the taper profile for a single, long helical undulator and the strongest focusing lattice with $L_{FODO}$ = 2.2 m with a parabolic transverse profile. All other beam and undulator parameters were identical to those used above for the Gaussian transverse profile; in particular, the rms beam radius used in the parabolic transverse profile is identical to that used for the Gaussian transverse profile.

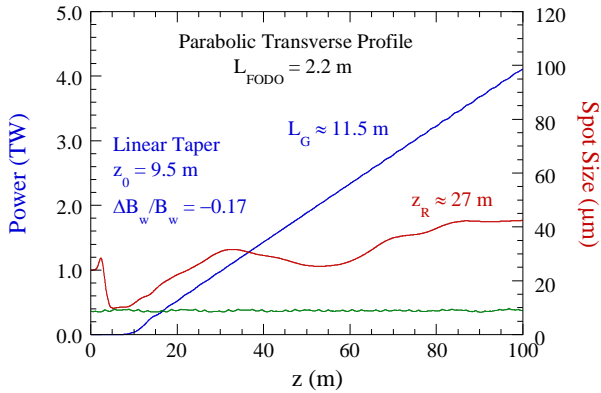

Fig. 12 Evolution of the power (left axis, blue) and spot size (right axis, red) for a parabolic transverse profile. The green line is the rms electron beam radius.

For this purpose, we have performed steady state simulations, and the evolution of the power and spot size of the optical field for the optimal taper profile is shown versus distance along the undulator in Fig. 12. The corresponding performance using a Gaussian transverse profile is shown in Fig. 3e where the output power reached about 2.7 TW. The optimal taper for the parabolic transverse profile starts after 9.5 m for a 5 MW seed with a total down taper at the 100 m point of 17%. The output power was found to be 4.1 TW which is substantially greater than that found using the Gaussian transverse profile. The characteristic growth length $L_G \approx$ 11.5 m which is longer than the Rayleigh range at the start taper point of 2.6 m. As a result, the optical field diffracts and the Rayleigh range increases to about 20 m at the 30 m point, exceeding the growth length, after which the optical field is largely guided.

## VI. SUMMARY AND DISCUSSION

The physics underlying efficiency enhancement in tapered undulators has been understood for decades [7,8]; however, tapered undulator experiments have historically shown enhancements in the efficiency over the saturated power in a uniform undulator of less than five [11-13]. In this paper, we have examined the effect of extreme focusing on the performance of tapered-undulator XFELs using the MINERVA simulation code and found that enhancements over the saturated power in a uniform undulator by a factor of 50 – 100 are possible. Optimizing the transverse profile, such as using a parabolic profile, may further enhance the output power. We considered resonant interactions at 1.5 Å with electron beams with energies and currents in the excess of 13 GeV and 4000 A respectively. The emittances we assumed were 0.3 – 0.4 mm-mrad with an rms energy spread of 0.01%. These parameters are consistent with what is achieved in the current range of XFELs.

The simulations indicate that the most important factor in achieving near-TW or TW power levels is the extreme focusing in a strong FODO lattice in conjunction with a long tapered undulator. This level of performance was found using either helical or planar undulators where the transverse focusing of the electron beam reached current densities in excess of 20 GA/cm$^2$. These extreme current densities are associated with large values for the Pierce parameter which give rise to extremely strong interactions that result in substantial optical guiding even in a tapered undulator configuration.

It is important to remark that the advantages accruing from such extreme focusing are found in preliminary time-dependent simulations for both self-seeded MOPAs and pure SASE XFELs. In particular, we find that the average power over the pulse achieved for the time-dependent simulation of the self-seeded MOPA is comparable to that found in the corresponding steady-state simulation. This indicates that slippage is not an important issue for the parameters under consideration. In addition, the time-dependent SASE simulation reached a power that is comparable to that of the self-seeded MOPA, indicating that TW power levels can also be achieved with pure SASE.

## ACKNOWLEDGMENTS

We would like to thank G.R. Neil for helpful discussions. This research used resources of the Argonne Leadership Computing Facility, which is a DOE Office of Science User Facility supported under Contract DE-AC02-06CH11357 for the preliminary time-dependent simulations.